\documentclass[showpacs,amssymb,aps,twocolumn]{revtex4}
\usepackage{amsmath}
\usepackage{graphicx}
\usepackage{epsf}
\usepackage{graphics}
\usepackage[latin1]{inputenc}
\newcommand{\be}{\begin{equation}}
\newcommand{\ee}{\end{equation}}
\newcommand{\bea}{\begin{eqnarray}}
\newcommand{\eea}{\end{eqnarray}}

\begin{document}

\title{Thermal instability in a gravity-like scalar theory}

\author{F. T. Brandt$^{a}$, Ashok Das$^{b,c}$ and  J. Frenkel$^{a}$}
\affiliation{$^{a}$ Instituto de Física, Universidade de São Paulo, 05508-090, São Paulo, SP, BRAZIL}
\affiliation{$^b$ Department of Physics and Astronomy, University of Rochester, Rochester, NY 14627-0171}
\affiliation{$^c$ Saha Institute of Nuclear Physics, 1/AF Bidhannagar, Calcutta 700064, India}

\begin{abstract}
We study the question of stability of the ground state of a scalar
theory which is a generalization of the $\phi^3$ theory
and has some similarity to gravity with a cosmological
constant. We show that the ground state of the theory at zero
temperature becomes unstable above a certain
critical temperature, which is evaluated in closed form at high temperature.
\end{abstract}

\pacs{04.60.-m,11.10.Wx}

\maketitle

\section{Introduction}
The purely attractive nature of gravity is a source of  instabilities
in a gravitational system. Nevertheless, in the presence of a positive
cosmological constant, such a system may remain in a static and
homogeneous state at zero temperature. It is interesting to inquire
whether this behavior may also hold at non-zero
temperature. Instabilities
in quantum gravity at finite temperature have been studied from various points of 
view
\cite{Gross:1982cv,kikuchi:1984np,Nakazawa:1984zq,gribosky:1989yk,Frenkel:1991dw,Rebhan:1991yr,Brandt:1992qn,Brandt:2008cn}.
These works indicate that, due to the inherent complexity of
Einstein`s theory, a complete analysis of
the stability problem at finite temperature becomes quite involved. For this reason, it
may be useful to study the stability issue in a simpler model, which
would mimic the behavior of gravity and yet would allow for an
all-order perturbative analysis of this problem. In
this spirit, such a study has already been undertaken by several
authors \cite{LeBellac:1989ps,Grandou:1990ir,Pisarski:1990ds,Altherr:1991fu}
within the context of the $\phi^3$ theory in six dimensions.

With a similar motivation, we consider in this note a non-polynomial generalization of
the $\phi^3$ theory with derivative interactions, which has several interesting analogies to gravity
with a cosmological constant. We analyze the effective
potential of this model, both at zero temperature as well as at finite
temperature. At zero temperature we find that this potential has a
unique minimum 
which indicates that the ground state of the system is stable.
We next carry out an analysis of the behavior of the effective
potential at finite temperature, and determine the value of the critical temperature
above which the system becomes unstable. The calculation is done in the tadpole
approximation, which provides the leading order contributions at high
temperature to all orders. In section {\bf II} we study the effective
potential at zero temperature and in section {\bf III} we extend this
analysis to the case of non-zero temperature. We conclude this note with 
a brief discussion in section {\bf IV}.

\section{The model and the effective potential at zero temperature}

Let us consider a scalar model, which is reminiscent of (scalar) 
gravity with a positive cosmological constant $\Lambda$, described by the
Lagrangian density
\be\label{1}
{\cal L} = \sqrt{1+\kappa \phi}
\left(\frac{1}{2}\partial^\mu\phi\partial_\mu\phi +
  \Lambda\right) + J \phi,
\ee
where we can think of $\phi$ as playing the role of the metric 
tensor and $J$ denotes an external source. The presence of
non-polynomial and  derivative interaction terms in this theory captures
some of the inherent 
characteristics of a theory of gravity. We may also think of this model as a generalization of the $\phi^3$
theory as follows. Let us expand \eqref{1} in a power series of
$\kappa\phi$ and express $\kappa$, $\Lambda$ and $J$ in terms of a set of new
parameters $g$, $m$ and source $j$ (we assume $g>0$) as
\be\label{2}
\kappa = - \frac{2}{3}\frac{g}{m^2}, \;\;  
\Lambda = \frac{9 m^6}{g^2},\,\; 
\,\; J = \frac{3 m^4}{g} + j.
\ee
%%%%%%%%%%%%%%%%%%%%%%%%%%%%%%%%%%%%%%%%%%%%%%%%%%%%%%%%%%%%%%%%%%%%%%%%%%%%%%%%
%                                        2  2           3  3            4  4
%                        g phi        phi  g         phi  g          phi  g
%    coeffKin := 1 + 1/3 ----- - 1/18 ------- + 1/54 ------- - 5/648 -------
%                           2             4              6               8
%                          m             m              m               m
%%%%%%%%%%%%%%%%%%%%%%%%%%%%%%%%%%%%%%%%%%%%%%%%%%%%%%%%%%%%%%%%%%%%%%%%%%%%%%%%
%              6                                     2    4
%             m          2    2            3        g  phi
%          9 ---- - 1/2 m  phi  + 1/6 g phi  - 5/72 ------- + 1/2 j phi
%              2                                        2
%             g                                        m
%%%%%%%%%%%%%%%%%%%%%%%%%%%%%%%%%%%%%%%%%%%%%%%%%%%%%%%%%%%%%%%%%%%%%%%%%%%%%%%%%
Then, we can write ${\cal L}$ in the form
\begin{eqnarray}\label{3}
{\cal L} & = & \sqrt{1 - \frac{2g}{3 m^{2}}\,\phi} 
\left(\frac{1}{2}\partial_{\mu}\phi \partial^{\mu}\phi 
+ \frac{9m^{6}}{g^{2}}\right) + \left(\frac{3m^{4}}{g} + j\right)\phi\nonumber\\
&=& \left(
1-\frac{1}{3}\frac{g}{m^2}\phi
-\frac{1}{18}\frac{g^2}{m^4}\phi^2
%+\frac{1}{54}\frac{g^3}{m^6}\phi^3
%\right. \nonumber \\ &-& \left.
%\frac{5}{648}\frac{g^4}{m^8}\phi^4
+\cdots
\right)\frac{1}{2}(\partial_\mu\phi)(\partial^\mu\phi)
\nonumber \\ 
&+& \frac{9m^{6}}{g^{2}} -\frac{m^2}{2}\phi^2-\frac{g}{3!}\phi^3
-\frac{5g^2}{(3m^2)4!}\phi^4+\cdots
+ j\phi .
\end{eqnarray}
This shows that the theory \eqref{1} can be interpreted as a
generalization of the massive $\phi^3$ theory
with nonpolynomial as well as derivative coupling interactions 
(much like in a theory of gravity)  in the presence of the
source $j$ for the scalar field. We also note here that the shift in the source in
\eqref{2} has been defined to eliminate, when $j$ vanishes, the linear terms in $\phi$ in
the Lagrangian density.
%(as is clear from \eqref{3}) so that $j$
%corresponds to the source for the scalar field (for the purpose of
%deriving amplitudes).

\begin{figure}[htbp]
\begin{center}
\begin{tabular}{c}
\includegraphics[scale=0.26]{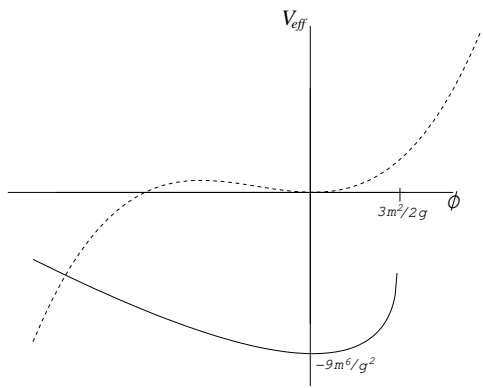}
\end{tabular}
\end{center}
\caption{The zero temperature effective potential for the $\phi^3$ theory (dashed line) and the theory
  described by Eq. \eqref{3} (solid line) for the same values of the
  parameters $m$ and $g$.}
\label{fig1}
\end{figure}

From \eqref{3}, we see that the theory is defined only for 
$\phi < 3m^{2}/2g$ 
and in this case, the Hamiltonian density of the theory (with $j=0$) is readily obtained to be
\be\label{4}
{\cal H} = \frac{1}{2}\sqrt{1- \frac{2g}{3m^{2}}\phi}
\left[\left(\dot\phi\right)^2+\left(\partial_i\phi\right)^2- 
\frac{18m^{6}}{g^{2}}\right]-\frac{3m^{4}}{g}\phi.
\ee
This expression shows that the system will attain a minimum energy for field 
configurations which are uniform in space-time. For such constant field configurations,
the stability of the vacuum is  conveniently studied through the effective 
potential \cite{Coleman:1973jx,dolan:1974qd,weinberg:1974hy,Fujimoto:1984yz,evans:1987ws},
which in our case has the simple form (we remark here that this is
really the tree level potential; the effective potential should also
contain loop corrections which we are neglecting with the assumption
that their effect is expected to be small)
\be\label{5}
V_{\mbox{\scriptsize{\it eff}}} = -\frac{9m^{6}}{g^{2}}\sqrt{1-\frac{2g}{3m^{2}}\phi} - \frac{3m^{4}}{g} \phi.
\ee
The unique extremum of this effective potential satisfying 
$\frac{dV_{\mbox{\scriptsize{\it eff}}}}{d\phi}=0$, occurs at 
\be\label{6}
\frac{1}{\sqrt{1- \frac{2g}{3m^{2}} \phi}}= 1,
\ee
which determines the vacuum expectation value of the scalar field in this theory to correspond to
\be\label{7}
\langle 0|\phi|0\rangle = \langle \phi\rangle = 0.
\ee
Furthermore, from Eq. \eqref{5}, we note that
\be\label{8}
\left.\frac{d^2 V_{\mbox{\scriptsize{\it eff}}}}{d\phi^2}\right|_{\phi=0} = m^{2} > 0,
\ee
so that the extremum of the potential is indeed a minimum. 
We note here that the local minimum in the $\phi^{3}$ theory also 
occurs at the origin. However, unlike the $\phi^{3}$ theory where the
potential is unbounded from below, here we have a potential that is
better behaved as shown in  Fig. \ref{fig1}. At zero temperature,
therefore, the vacuum of the theory is stable.

\section{Stability of the ground state at high temperature}

Let us next study the effective potential for the theory in \eqref{3} in the presence of a heat
bath at temperature $T$. This effective potential can be calculated using either
the real time formalism or the imaginary time formalism %in a straightforward manner
\cite{kapusta:book89,lebellac:book96,das:book97} and will contain
thermal contributions of various kinds in addition to the zero
temperature contributions in \eqref{5}. However, using a similar analysis to that presented in reference
\cite{Altherr:1991fu}, it may be shown that at high temperature $T\gg m$, the
leading thermal contribution to the effective potential 
arises only from the tadpole (one-point) diagram. In this
regime, the thermal contribution of the tadpole at one-loop can be
%easily 
calculated in a straightforward manner from the thermal Feynman rules following from the
Lagrangian density in \eqref{3}. 
%and in $n$-dimensional space-time has the value
The momentum integration can be done using dimensional regularization
and the relation (for even values of $n$) \cite{gradshteyn} 
\be
\int_0^\infty dk k^{n-3} N_B(k) = (-1)^{n/2} (2\pi T)^{n-2}\frac{B_{n-2}}{2(n-2)},
\ee
where $N_B(k)=({\rm e}^{k/T}-1)^{-1}$ is the bosonic distribution
function and $B_{n-2}$ are the Bernoulli numbers. This leads in an
even $n$-dimensional space-time to the leading temperature-dependent
part of the tadpole
\be\label{9}
\Gamma_{\mbox{\scriptsize{\rm tad}}}^{(T)} = g
\frac{\pi^{(n-3)/2}}{3(n-2)}\frac{|B_{n-2}|}{\Gamma\left(\frac{n-1}{2}\right)}
T^{n-2}\equiv  g C_{n} T^{n-2}.
\ee
where $\Gamma$ is the gamma
function \cite{gradshteyn} 
and we have identified
\be\label{10}
C_{n} = \frac{\pi^{(n-3)/2}}{3(n-2)}\frac{|B_{n-2}|}{\Gamma\left(\frac{n-1}{2}\right)}.
\ee

\begin{figure}[htbp]
\begin{center}
\begin{tabular}{c}
\includegraphics[scale=0.26]{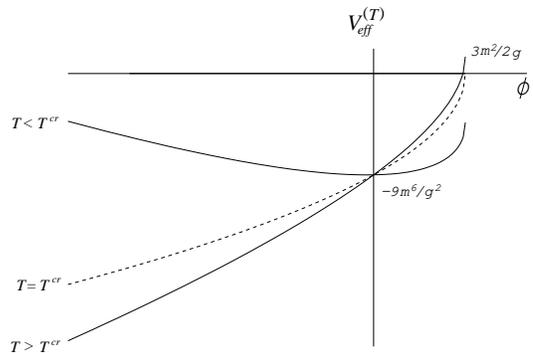}
\end{tabular}
\end{center}
\caption{The thermal effective potential for the theory in \eqref{3}.}
\label{fig2}
\end{figure}

As a result, the leading behavior of the effective potential at high temperature is given by
\be\label{11}
V_{\mbox{\scriptsize{\it eff}}}^{(T)} = V_{\mbox{\scriptsize{\it eff}}} + \Gamma_{\mbox{\scriptsize{\rm tad}}}^{(T)} \phi = V_{\mbox{\scriptsize{\it eff}}} + g C_{n} T^{n-2} \phi,
\ee
where $V_{\mbox{\scriptsize{\it eff}}}$ is the zero temperature effective potential defined in \eqref{5}. The extremum of the effective potential now satisfies
\begin{eqnarray}\label{12}
\frac{1}{\sqrt{1 - \frac{2g}{3m^{2}} \phi}} &=& 1 - \frac{g}{3m^{4}}\Gamma_{\rm tad}^{(T)}
\nonumber \\
                                           &=& 1 - \frac{g^{2}}{3m^{4}} C_{n} T^{n-2},
\end{eqnarray}
leading to a temperature dependent vacuum expectation value
\begin{eqnarray}\label{13}
\langle\phi\rangle^{(T)} &=& \frac{3m^{2}}{2g}
\left[1 - \left(1 - \frac{g}{3m^{4}} \Gamma_{\rm tad}^{(T)}\right)^{-2}\right]
\nonumber \\
&=& \frac{3m^{2}}{2g}
\left[1 - \left(1 - \frac{g^{2}}{3m^{4}} C_{n} T^{n-2}\right)^{-2}\right].
\end{eqnarray}
Furthermore, at the extremum we have
\be\label{14}
\frac{d^{2}V_{\mbox{\scriptsize{\it eff}}}^{(T)}}{d\phi^{2}}
\bigg|_{\langle\phi\rangle^{(T)}} = m^{2}\left(1 - \frac{g^{2}}{3m^{4}} C_{n} T^{n-2}\right)^{3}.
\ee
Thus, the second derivative in \eqref{14} would be positive and the
extremum will indeed correspond to a minimum as long as the
temperature is smaller than
\be\label{15}
T^{cr}_{(n)} = \left(\frac{3m^{4}}{g^{2} C_{n}}\right)^{\frac{1}{n-2}}.
\ee
This, therefore, defines a critical temperature for the system (in $n$ dimensions)  
and for temperatures below this critical temperature the ground state
will be stable. We see from Fig. \ref{fig2} that for large negative
values of $\phi$, the temperature dependent tadpole contribution has 
the effect of lowering the potential such that for temperatures below 
the critical temperature, there is a barrier. 
However, for temperatures (equal to or) above the critical temperature,
there is no longer a barrier and the thermal fluctuations induce a roll
off to infinity as shown in Fig. \ref{fig2}.

At this point, it may be instructive to give a diagrammatic
representation of the solution  which minimizes the
effective potential and determines the vacuum
expectation value of the field. Expanding out the effective potential in \eqref{11} we note that we can write
\be\label{17}
V_{\mbox{\scriptsize{\it eff}}}^{(T)} = - \frac{9m^{6}}{g^{2}} + \frac{m^2}{2}\phi^{2}+\frac{g}{3!}\phi^3 
+\frac{1}{4!}\frac{5g^{2}}{(3m^{2})}\phi^4+\cdots + \Gamma_{\mbox{\scriptsize{\rm tad}}}^{(T)} \phi.
\ee
Furthermore, recognizing that for constant field configurations 
we can represent $G^{-1} = -(\Box+m^{2})\rightarrow - m^{2}$ 
we can write the equation for the minimum of the effective potential also as
\begin{eqnarray}\label{18}
G^{-1}\phi &= &  
\Gamma_{\mbox{\scriptsize{\rm tad}}}^{(T)} 
+ \frac{g}{2!} \phi^{2} + \frac{1}{3!}\frac{5g^{2}}{3 m^2} \phi^{3} + \cdots ,
{\mbox or,}\;\;
\nonumber\\
\phi & = & G\left(\Gamma_{\mbox{\scriptsize{\rm tad}}}^{(T)} 
+ \frac{g}{2!} \phi^{2} + \frac{1}{3!}\frac{5g^{2}}{3 m^2} \phi^{3} + \cdots\right),
\end{eqnarray}
where $G=-1/m^2$ represents the scalar propagator evaluated at zero
momentum. Iterating this equation we obtain a
perturbative solution of Eq. \eqref{18} as a power series in $\Gamma_{\mbox{\scriptsize{\rm tad}}}^{(T)}$,
which can be represented graphically as
\begin{eqnarray}\label{19}
\langle \phi \rangle 
&=&
\begin{array}{c}\includegraphics[scale=0.2]{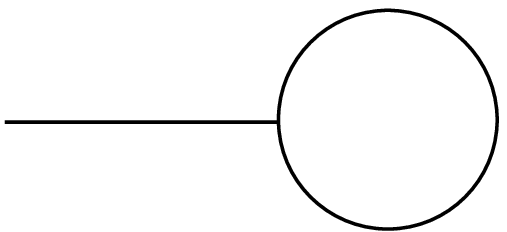}\end{array}
+\frac{1}{2} \begin{array}{c}\includegraphics[scale=0.2]{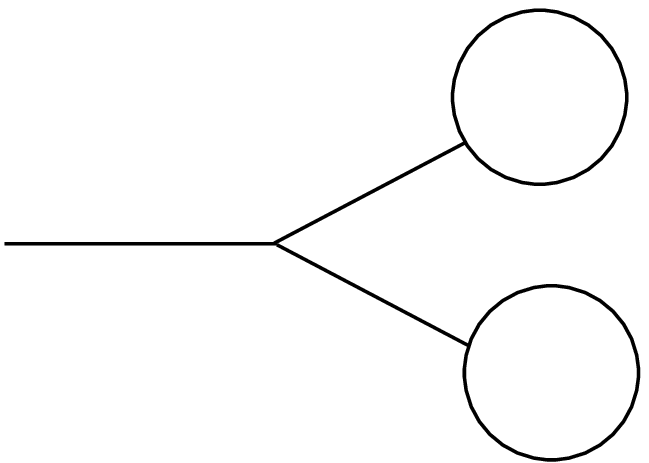}\end{array}
\nonumber \\
&+&\frac{1}{2} \begin{array}{c}\includegraphics[scale=0.2]{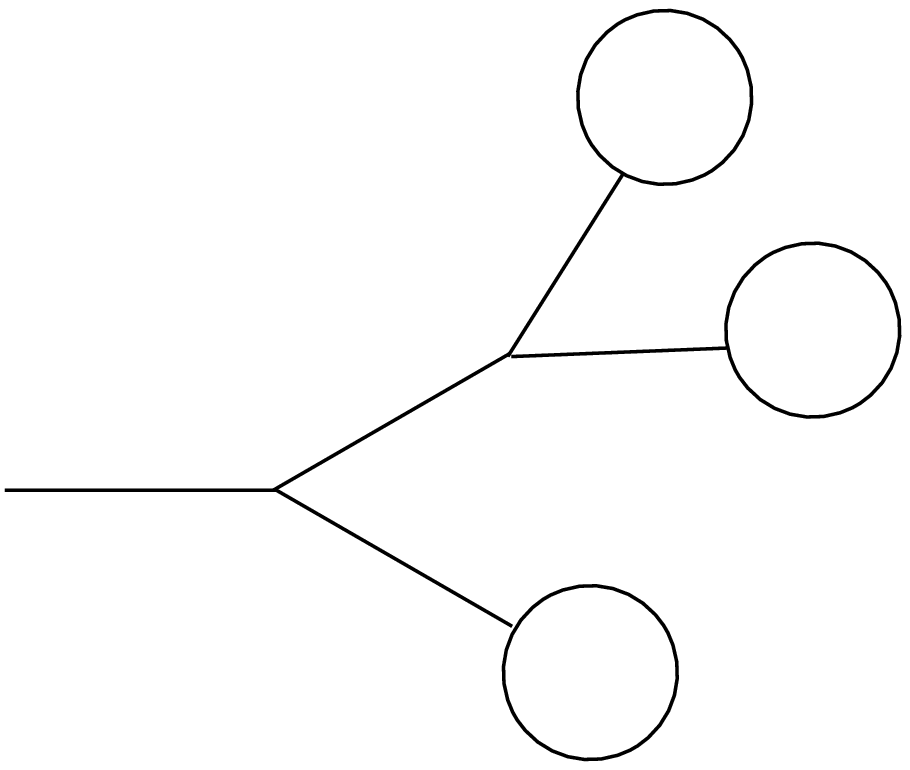}\end{array}
+\frac{1}{3!} \begin{array}{c}\includegraphics[scale=0.2]{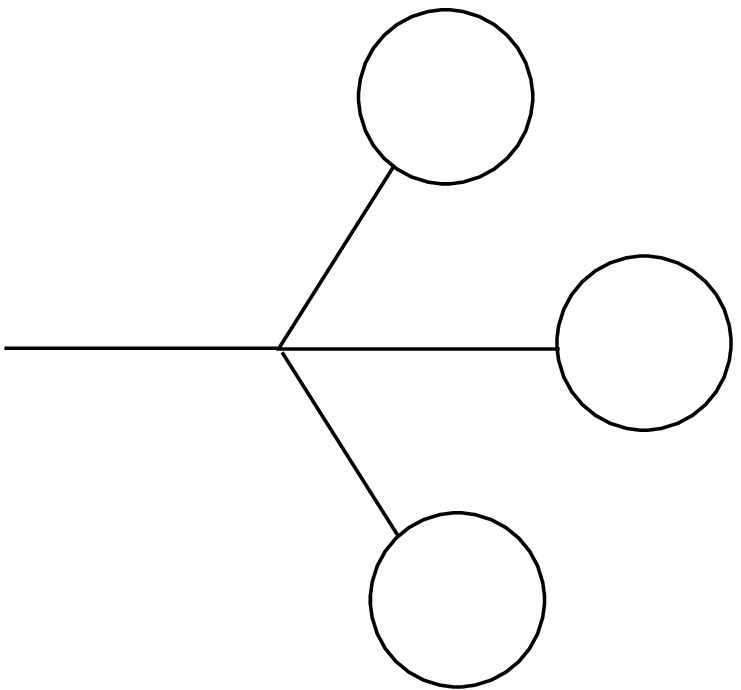}\end{array}
\nonumber \\
&+&\frac{1}{8} \begin{array}{c}\includegraphics[scale=0.2]{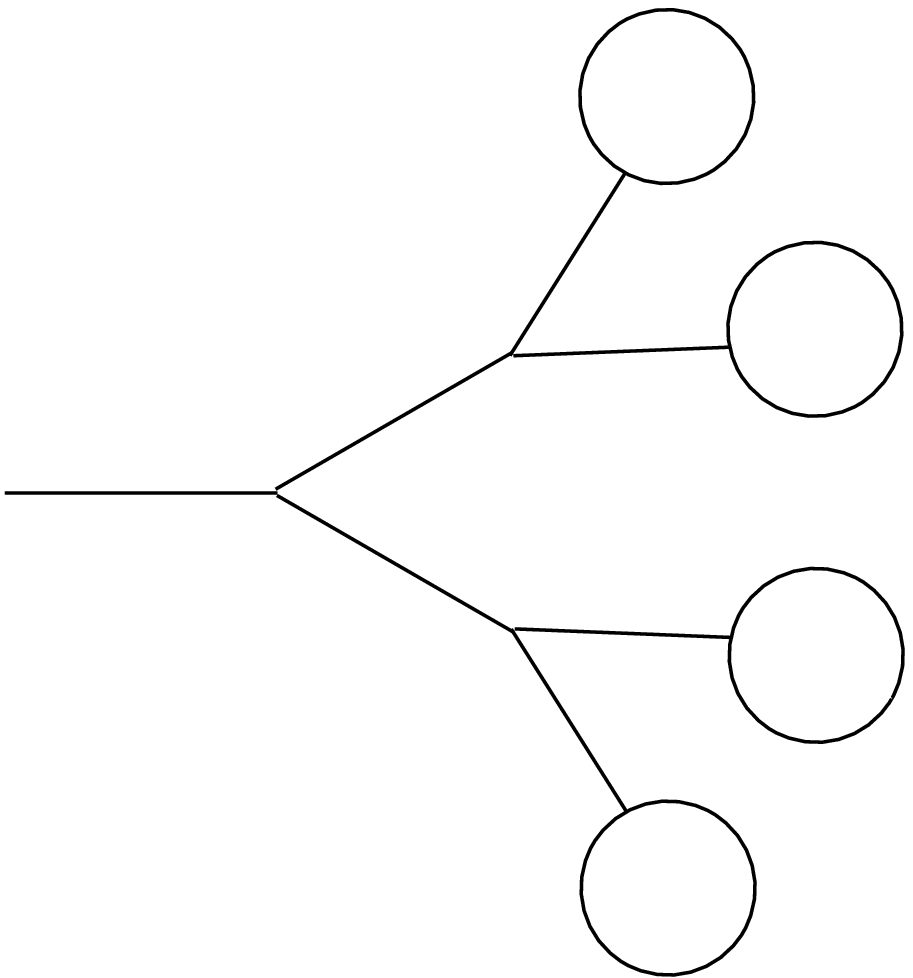}\end{array}
+\frac{1}{4} \begin{array}{c}\includegraphics[scale=0.2]{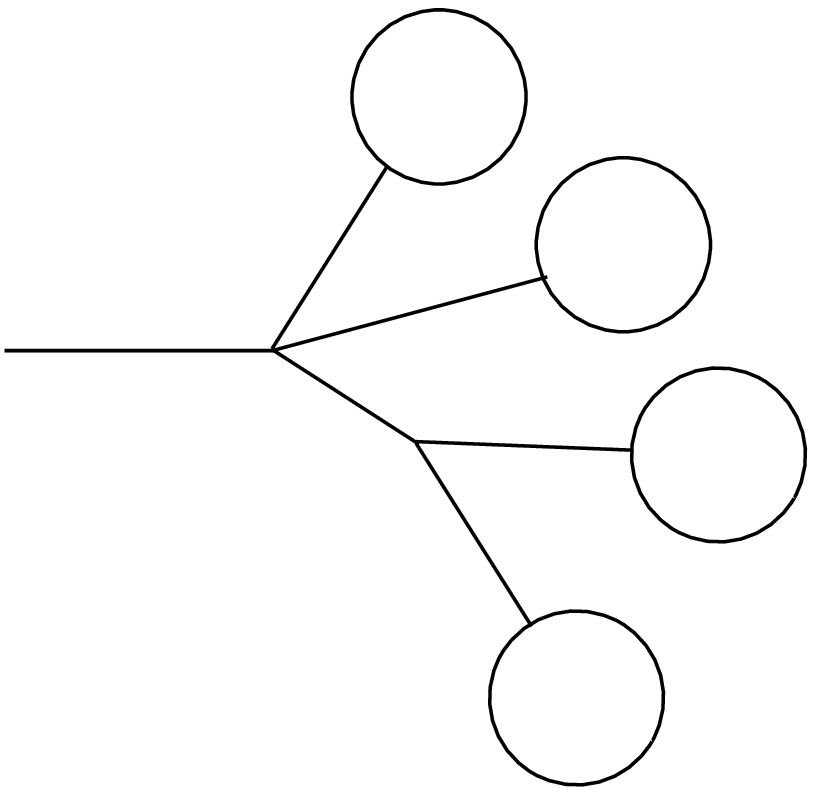}\end{array}
\nonumber \\
&+&\frac{1}{4!}\begin{array}{c}\includegraphics[scale=0.2]{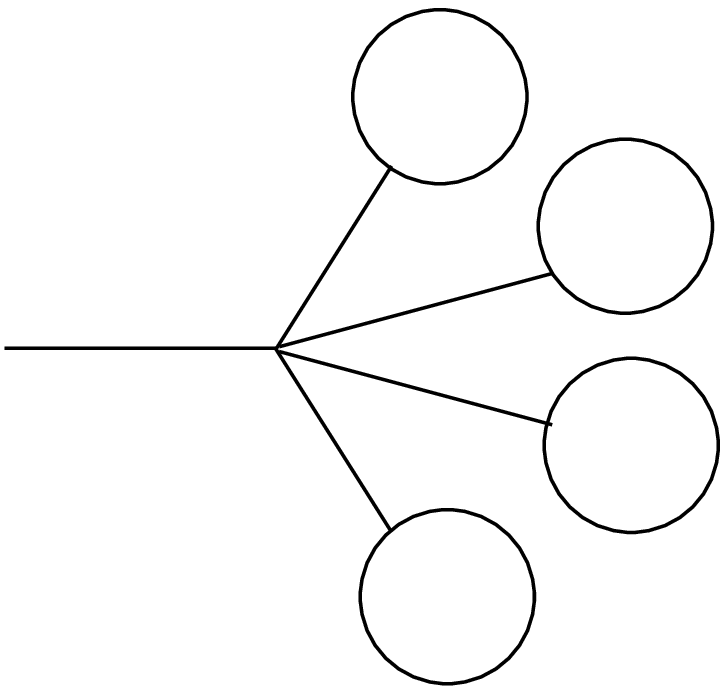}\end{array}
+\frac{1}{2} \begin{array}{c}\includegraphics[scale=0.2]{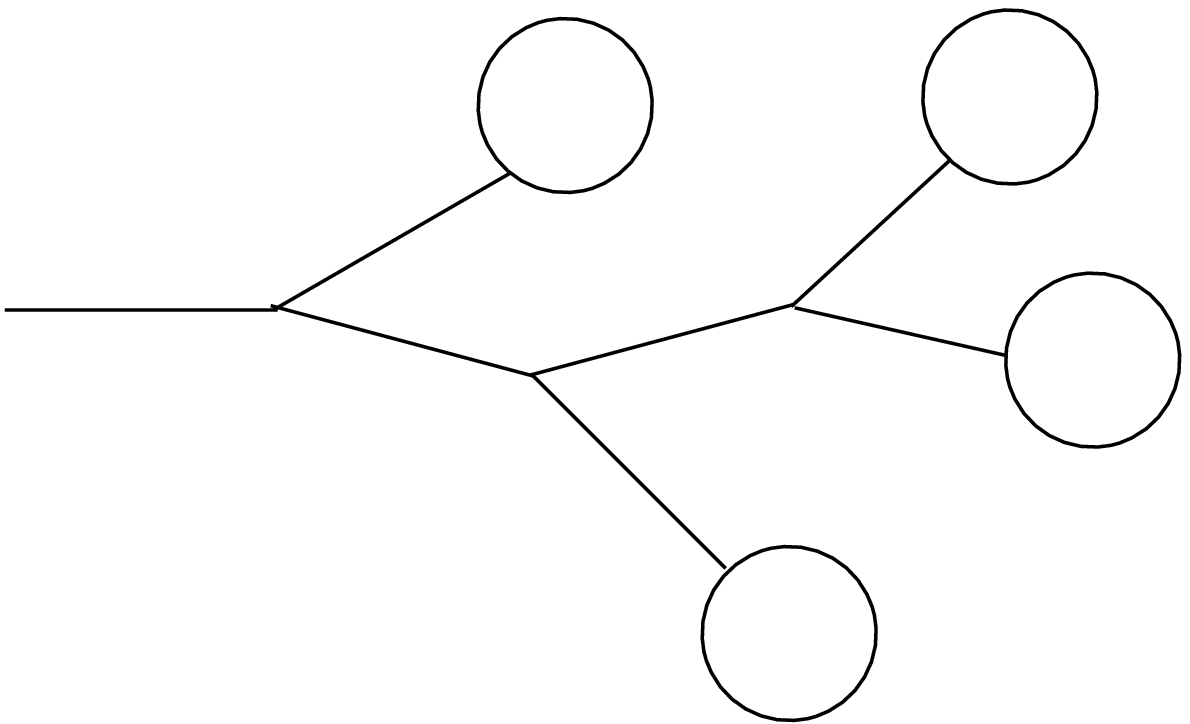}\end{array}
\nonumber \\
&+&\frac{1}{3!} \begin{array}{c}\includegraphics[scale=0.2]{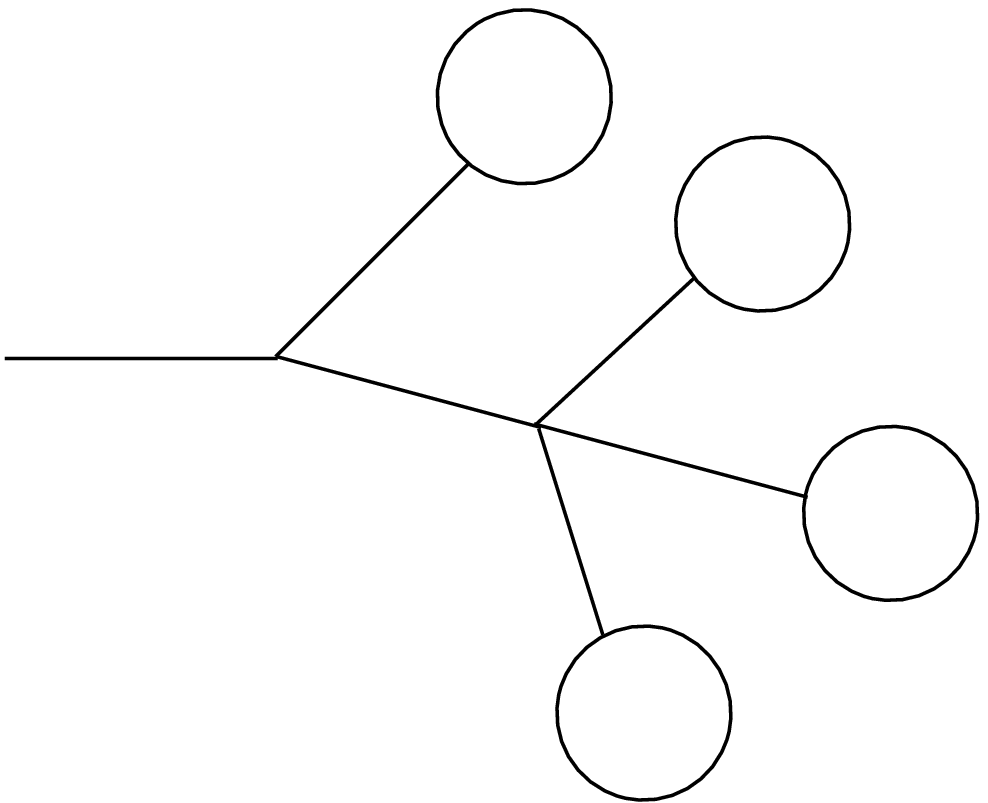}\end{array}
+\cdots .
\end{eqnarray}
Here the lines denote the zero-momentum (and zero temperature) propagator $G$, the blobs
represent the temperature-dependent tadpole $\Gamma_{\mbox{\scriptsize{\rm tad}}}^{(T)}$,
defined in \eqref{9},
and the vertices are: $\lambda_3=g$, $\lambda_4={5g^2}/{(3m^2)}$,
$\lambda_5={35g^3}/{(9m^4)}$, $\cdots$.
It can be checked that the sum of the above series of diagrams 
yields identically the result in \eqref{13}.

\section{Discussion}
We have studied the question of stability of the ground state of a scalar theory which may
be thought as a generalization of the conventional $\phi^3$ theory 
with nonpolynomial and derivative interactions. In this model,
which is somewhat analogous to gravity with a cosmological constant,
the critical temperature above which the theory becomes unstable is
given by Eq. \eqref{15}, which is our main result.

Let us now evaluate this temperature in six space-time dimensions, in
order to compare it with the critical temperature which occurs in the
renormalizable
$(\phi^3)_{6}$ model \cite{Altherr:1991fu}. Using the value of $C_n$ given in
Eq. \eqref{10}, we find that
\be\label{20}
T^{cr}_{(6)} = \left(\frac{810}{\pi} %\frac{9}{2}\frac{180}
\right)^{1/4} \frac{m}{\sqrt{g}}.
\ee
This critical temperature is somewhat higher than the one found in the
$(\phi^3)_6$ model, which indicates that the present theory may be
a bit more stable under thermal fluctuations.
          
In conclusion, let us point out an important difference between these
two theories. The model described by the Lagrangian density \eqref{3}
involves effective dimensionfull coupling constants, with dimension of
inverse powers of mass, and is, 
like gravity, non-renormalizable.
It turns out that, in order to be
able to neglect higher order thermal loops, it is necessary to assume
the condition
\be\label{21}
\frac{g^2 T^{n-4}}{m^2} = \left(\frac{g^2 T^{n-6}}{m^2}\right) T^2 \ll 1.
\ee
This relation is somewhat similar to the one required in four
dimensional thermal gravity \cite{Rebhan:1991yr}:
$G_{N} T^2\ll 1$, where $G_{N}$ is the gravitational constant. We remark
that Eq. \eqref{21} is consistent with the result \eqref{15} for the
critical temperature, provided the coupling constant $g$ is
sufficiently small.
For completness we note here that all of our results reduce,
in six space-time dimensions, to those
of \cite{Altherr:1991fu} upon proper truncation.

\bigskip

\noindent{\bf Acknowledgments}

We would like to thank Professor J. C. Taylor for
helpful discussions. This work was supported in part 
by US DOE Grant number DE-FG 02-91ER40685. 
We are grateful to CNPq and FAPESP, Brazil, for financial support.

%\bibliographystyle{prsty}
%\bibliography{all_new}

\end{document}